\documentclass[10pt,twocolumn,letterpaper]{article}

\usepackage{cvpr}
\usepackage{times}
\usepackage{epsfig}
\usepackage{graphicx}
\usepackage{amsmath}
\usepackage{amssymb}
\usepackage{multirow}

\usepackage[breaklinks=true,bookmarks=false]{hyperref}

\cvprfinalcopy 

\def\cvprPaperID{****} 

\ifcvprfinal\fi
\begin{document}

\title{Learning Deep Image Priors for Blind Image Denoising}

\author{Xianxu Hou $^{1}$ \quad Hongming Luo $^{1}$ \quad Jingxin Liu $^{1}$  \quad Bolei Xu $^{1}$  \\ \quad Ke Sun $^{2}$  \quad Yuanhao Gong $^{1}$   \quad Bozhi Liu $^{1}$ \quad Guoping Qiu $^{1,3}$\\
$^1$ College of Information Engineering and Guangdong Key Lab for Intelligent Information Processing,\\
Shenzhen University, China \\ $^2$  School of Computer Science, The University of Nottingham Ningbo China \\ $^3$ School of Computer Science, The University of Nottingham, UK\\
}

\maketitle
\thispagestyle{empty}

\begin{abstract}
Image denoising is the process of removing noise from noisy images, which is an image domain transferring task, \ie, from a single or several noise level domains to a photo-realistic domain. In this paper, we propose an effective image denoising method by learning two image priors from the perspective of domain alignment. We tackle the domain alignment on two levels. 1) the feature-level prior is to learn domain-invariant features for corrupted images with different level noise; 2) the pixel-level prior is used to push the denoised images to the natural image manifold. The two image priors are based on $\mathcal{H}$-divergence theory and implemented by learning classifiers in adversarial training manners. We evaluate our approach on multiple datasets. The results demonstrate the effectiveness of our approach for robust image denoising on both synthetic and real-world noisy images. Furthermore, we show that the feature-level prior is capable of alleviating the discrepancy between different level noise. It can be used to improve the blind denoising performance in terms of distortion measures (PSNR and SSIM), while pixel-level prior can effectively improve the perceptual quality to ensure the realistic outputs, which is further validated by subjective evaluation.
\end{abstract}

\section{Introduction}
Image denoising is a fundamental problem in low-level vision as well as an important pre-processing step for many other image restoration problems \cite{Zhang_2017_CVPR,heide2014flexisp}. It aims at recovering a noise-free image $x$ from its noisy observation(s) $y$ by following the degradation model $y = x + v$. As in many previous literatures \cite{buades2005non,Dong2013NonlocallyCS,Zhang_2017_CVPR,mairal2009non,zhang2018ffdnet}, $v$ is usually assumed additive white Gaussian noise (AWGN) of standard deviation $\sigma$. Therefore, prior knowledge modeling on images plays an essential part in image denoising.

The main success of the recent image denoising methods comes from the effective image prior modeling over the input images \cite{buades2005non,Dong2013NonlocallyCS,gu2014weighted}. State of the art model-based methods such as BM3D \cite{dabov2007image} and WNNM \cite{gu2014weighted} can be further extended to remove unknown noises. However, there are a few drawbacks of these methods. First, these methods usually involve a complex and time-consuming optimization process in the testing stage. Second, the image priors employed in most of these approaches are hand-crafted, such as nonlocal self-similarity and gradients, which are mainly based on the internal information of the input image without any external information.

In parallel, there is another type of denoising methods based on discriminative learning. They aim to train a deep denoising network with paired training datasets (noisy and clear images) and learn the underlying noise model implicitly to achieve fast inference \cite{burger2012image,Zhang_2017_CVPR,zhang2018ffdnet,jain2009natural}, among which DnCNN \cite{zhang2017beyond} and FFDNet \cite{zhang2018ffdnet} have obtained remarkable results. However existing discriminative learning methods are usually designed to a specific noise level with limited flexibility. Though DnCNN-B \cite{zhang2017beyond} can be trained for different noise levels for blind image denoising, it still cannot generalize well to real-world noisy images. FFDNet \cite{zhang2018ffdnet} still requires a tunable noise level map as the input to tackle various noise levels. Therefore, it is of great interest to develop general image priors which can help handle image denoising with a wide range of noise levels and generalize well for real-world noisy images.

To this end, we propose a new image denoising model, referred to Deep Image Prior Network (DIPNet), based on data-driven image priors. In particular, we consider image denoising as a domain transferring problem, \ie, from noise domain to photo-realistic domain. Inspired by this, we propose two image priors: 1) the feature-level prior which is designed to help decrease domain discrepancy between corrupted images with different noise levels for robust image denoising; 2) the pixel-level prior which is used to push the denoised image to photo-realistic domain for perceptual improvement. In particular, we model both priors as discriminator networks, which are trained by an adversarial training strategy to minimize the $\mathcal{H}$-divergence between different image domains.


The contribution of this work can be summarized as follows:

\begin{itemize}
   \item We propose an effective deep residual model based on data-driven image priors, namely DIPNet, for blind image denoising. Our method can achieve state of the art results for both synthetic and real-world noise removal.

   \item We design two image priors based on adversarial training. The feature-level prior is capable of alleviating the discrepancy between different noise levels to improve the denoising performance, while pixel-level prior can effectively improve the perceptual quality and produce photo-realistic results.

   \item Compared with previous methods, our method significantly improves the generalizability when adapting from synthetic Gaussian denoising to real-world noise removal. In particular, a single model trained for blind Gaussian noise removal can outperform competing methods designed specifically for real-world noise removal.

\end{itemize}
\section{Related Work}
\subsection{Image Denoising}
A large number of image denoising methods have been proposed over the recent years, and generally they can be grouped into two major categories: model-based methods and discriminative learning based methods.

\textbf{Model-based Methods} are usually depended on human-crafted image priors such as nonlocal self-similarity \cite{Dong2013NonlocallyCS,mairal2009non,buades2005non}, sparsity \cite{elad2006image} and gradients \cite{rudin1992nonlinear,weiss2007makes}. Two of the classic methods are BM3D \cite{dabov2007image} and WNNM \cite{gu2014weighted}, which are usually used as the benchmark methods for image denoising. In particular, BM3D uses an enhanced sparse representation for denoising by grouping similar 2D image fragments into 3D data arrays. WNNM proposes to use weighted nuclear norm minimization for image denoising by exploiting image nonlocal self-similarity. These models can be also further extended to handle blind denoising problem with various noise levels. In addition, a few approaches \cite{liu2008automatic,rabie2005robust,zhu2016noise} are also proposed to directly address blind image denoising problem by modeling image noise to assist corresponding denoising algorithms. However, these models are based on human-crafted priors which are designed under limited observations. In this paper, instead of only using the internal information of the input images, we propose to automatically learn image priors by making full use of external information.

\textbf{Discriminative Learning Based Methods} try to model image prior implicitly with paired (noisy and clear images) training data. These models have achieved great success in image denoising by taking advantage of the recent development of deep learning. Several approaches adopt either a plain Multilayer Perceptron (MLP) \cite{burger2012image} or convolutional neural networks (CNN) \cite{jain2009natural,xie2012image,agostinelli2013adaptive,lefkimmiatis2017non,Zhang_2017_CVPR,chen2018deep} to learn a non-linear mapping from noisy images to photo-realistic ones. It is worth to mention that remarkable results have been obtained by recent deep learning based models. DnCNN \cite{zhang2017beyond} successfully trains a deep CNN model with batch normalization and residual learning to further boost denoising performance.
Moreover, DnCNN can be extended to handle noisy images with different level noise. A generic image-to-image regression deep model (RBDN) \cite{santhanam2017generalized} can be effectively used for image denoising. A deep kernel prediction network \cite{mildenhall2018burst} is trained for denoising bursts of images taken from a handheld camera. GAN-CNN \cite{chen2018image} proposes to use generative adversarial network (GAN) to estimate the noise distribution for blind image denoising. Furthermore, FFDNet \cite{zhang2018ffdnet} presents a fast and flexible denoising convolutional neural network, which can handle a wide range of noise levels with a tunable noise level map. Recently deep image prior \cite{ulyanov2018deep} is also proposed for general image restoration. However, most previous discriminative learning based methods have to learn multiple models for handling images with different noise levels. It is still a challenging issue to develop a single discriminative model for general image denoising for both synthetic and real-world noises.

\textbf{Deep Learning on Image Transformation.} Beyond image denoising, deep CNNs have been successfully applied to other image transformation tasks, where a model receives a certain input image and transforms it into the desired output. These applications include image super-resolution \cite{dong2016image}, downsampling\cite{hou2017deep}, colorization \cite{larsson2016learning}, deblurring \cite{li2018learning}, style transfer \cite{johnson2016perceptual}, semantic segmentation \cite{long2015fully}, image synthesis \cite{hou2017deepvae,goodfellow2014generative} \etc. In addition, \cite{blau2018perception} analyses the trade-off between the distortion and perception measures for image restoration algorithms. However their models are designed to handle input images within a specific domain, and cannot be directly used for blind image denoising with unknown noise level. In this work, we focus on using a single model to effectively handle a wide range of noise levels with data-driven image priors and also investigate the perception-distortion trade-off in terms of image denoising.

\section{Method}
Our goal is to develop a model which can take the noisy images to produce photo-realistic ones. The primary challenges are twofold: first, our model should be flexible and robust to process the same images corrupted with different level noise; second, we must ensure that the denoised images are realistic and visually pleasing. To address these challenges, we propose to learn two image priors based on $\mathcal{H}$-divergence theory, considering that the corrupted images with different level noise as well as clear images are in different image domains.

\subsection{Distribution Alignment with $\mathcal{H}$-divergence}
\label{sec:hdivergence}
The $\mathcal{H}$-divergence \cite{ben2010theory,ben2007analysis} is proposed to estimate the domain divergence from two set of unlabeled data with different distributions. It is a classifier-induced divergence through learning a hypothesis from a class of finite complexity. For simplicity, we first consider the $\mathcal{H}$-divergence of two set of samples. Thus, the problem of distribution alignment of the two domains can be formulated through a binary classification. Specifically we define a domain as a distribution $\mathcal{D}$ on inputs $\mathcal{X}$, thus $x'$ and $x''$ can be denoted as samples belonging to the two different domains $\mathcal{D'}$ and $\mathcal{D''}$ respectively. We also denote a labeling function $h: \mathcal{X} \rightarrow [0, 1] $ as a domain classifier in order to predict different labels of samples to 0 or 1 for domain $\mathcal{D'}$ and $\mathcal{D''}$ respectively. Let $\mathcal{H}$ be a hypothesis class on $\mathcal{X}$, \ie, a set of possible domain classifiers $h \in \mathcal{H}$. The $\mathcal{H}$-divergence between domain $\mathcal{D'}$ and $\mathcal{D''}$ is defined as follows:

\begin{scriptsize}
\begin{equation}
d_{\mathcal{H}}(\mathcal{D'}, \mathcal{D''}) = 2 \Big(1 - \min_{h \in \mathcal{H}} \Big[ \frac{1}{N} \sum_x \mathcal{L}_{\mathcal{D'}} (h(x))+ \frac{1}{N} \sum_{x}\mathcal{L}_{\mathcal{D''}}(h(x))\Big]\Big)
\end{equation}
\end{scriptsize}
where $\mathcal{L}_{\mathcal{D'}}$ and $\mathcal{L}_{\mathcal{D''}}$ denote the loss of label prediction $h(x')$ and $h(x'')$ on domain $\mathcal{D'}$ and $\mathcal{D''}$ respectively. $N$ is the total number of samples for a given dataset. We can see that the domain distance $\mathcal{H}$ is inversely proportional to the loss of the optimal domain classifier $h(x)$.

Under the context of deep learning, $x$ can be defined as the output image or hidden activations produced by a neural network $f$. Therefore, in order to reduce the dissimilarity of the distributions $\mathcal{D}$ and $\mathcal{D'}$, we can train $f$ to maximize the loss of the domain classifier. As a result, we need to play a maxmin game between $f$ and $h$ as follows:

\begin{scriptsize}
\begin{equation}
\min_f{d_{\mathcal{H}}(\mathcal{D'}, \mathcal{D''})} \Leftrightarrow \max_f \min_{h \in \mathcal{H}} \{ \frac{1}{N} \sum_x \mathcal{L}_{\mathcal{D'}} (h(x))+ \frac{1}{N} \sum_{x}\mathcal{L}_{\mathcal{D''}}(h(x))\}
\end{equation}
\end{scriptsize}
Furthermore, when considering multiple sets of samples with $m$ domains denoted as $\mathcal{D}^1, \mathcal{D}^2 ..., \mathcal{D}^m$. The distribution alignment can be formulated to a similar image optimization problem as follows:

\begin{scriptsize}
\begin{equation}
\begin{split}
\min_f{d_{\mathcal{H}}(\mathcal{D}^1, \mathcal{D}^2, ..., \mathcal{D}^m)} &\Leftrightarrow \max_f \min_{h \in \mathcal{H}} \{ \frac{1}{N} \sum_{m} \sum_{x} \mathcal{L}_{\mathcal{D}^m} (h(x))\}
\end{split}
\end{equation}
\end{scriptsize}
In practice, this optimization can be achieved in an adversarial training manner by two ways. One is to adopt generative adversarial networks (GANs) framework \cite{goodfellow2014generative} by reversing the label of the two categories; the other is to integrate a gradient reversal layer (GRL) \cite{ganin2015unsupervised,chen2018domain} in CNN model. In particular, GRL is designed to retain the input unchanged during forward propagation and reverses the gradient by multiplying it with a negative constant during the backpropagation. Additionally, GRL can be easily extended to multi-class inputs while GAN is more appropriate to deal with inputs with two categories. In our work, GRL and GAN are used to train feature-level and pixel-level priors respectively.

\subsection{Learnable Image Priors}
\label{sec:image_prior}
\textbf{Motivation.}
Most classic image denoising models can be formulated to solve the following problem \cite{zhang2018ffdnet}:
\begin{equation}
\label{eq:minimize}
\hat{x} = \arg \min_x \frac{1}{2\sigma} \|y-x\|^2 + \lambda P(x)
\end{equation}
where the first part $\frac{1}{2\sigma} \|y-x\|^2$ is the data fidelity term with different noise level $\sigma$, the second part $P(x)$ is the regularization term with image prior which is usually predefined. $\lambda$ is the hyper-parameter to balance the two parts. A discriminative denoising model, which is adopted in this work, aims to learn a non-linear mapping function $x = \mathcal{F}(y)$ parameterized by $W$ to predict the latent clear image $x$ from noisy image $y$. Thus, the solution of Equation \ref{eq:minimize} is given by:
\begin{equation}
\label{eq:no_prior}
\hat{x} = \mathcal{F}(y, \sigma, \lambda, P; W)
\end{equation}
The key to the success of this framework lies on the predefined image prior. This observation motivates us to learn image priors directly from image data. In particular, we propose to learn two data-driven image priors on feature level and pixel level respectively.

\begin{figure}[!tb]
\begin{center}
\includegraphics[width=8cm]{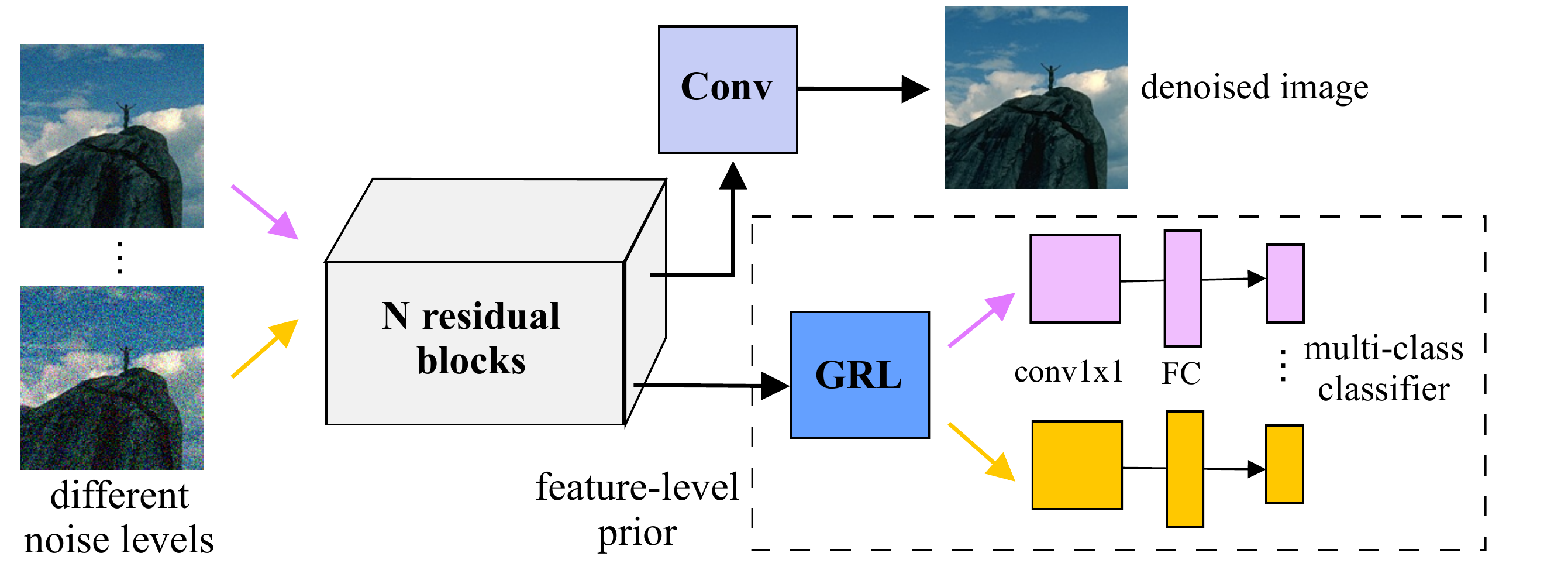}
\end{center}
   \caption{The dashed rectangle is used to construct the loss function to learn feature-level prior.
   }
\label{fig:feat_prior}
\end{figure}

\textbf{Feature-level Prior.}
Equation \ref{eq:no_prior} requires predefined noise level $\sigma$, thus the trained model cannot be flexible enough to handle different noise levels with a single network. In order to achieve blind image denoising, we seek to incorporate noise level information by learning an image prior in the feature space. In particular, we train a multi-class discriminator on the output of fused features from local and global path (see Section \ref{sec:network}) for different noise level images as shown in Figure \ref{fig:feat_prior}.
We try to learn the feature-level prior ($P_{feat}$) via a multi-class cross entropy loss as follows:
\begin{equation}
\label{eq:prior}
\mathcal{L}_{p_{feat}} = - \frac{1}{N} \sum_{i=1}^N log\Big(\frac{e^{\hat{p_i}}}{\sum_{j=1}^m{e^{p_{i,j}}}}\Big)
\end{equation}
Here $m$ denotes the number of noise levels. $p_{i, j}$ is the output score for $j^{th}$ class for a given image $i$ and $\hat{p_i}$ is output score for the correct class. We add a gradient reversal layer (GRL) before the multi-class discriminator to achieve the adversarial training as shown in Figure \ref{fig:feat_prior}.

\begin{figure}[!tb]
\begin{center}
\includegraphics[width=8cm]{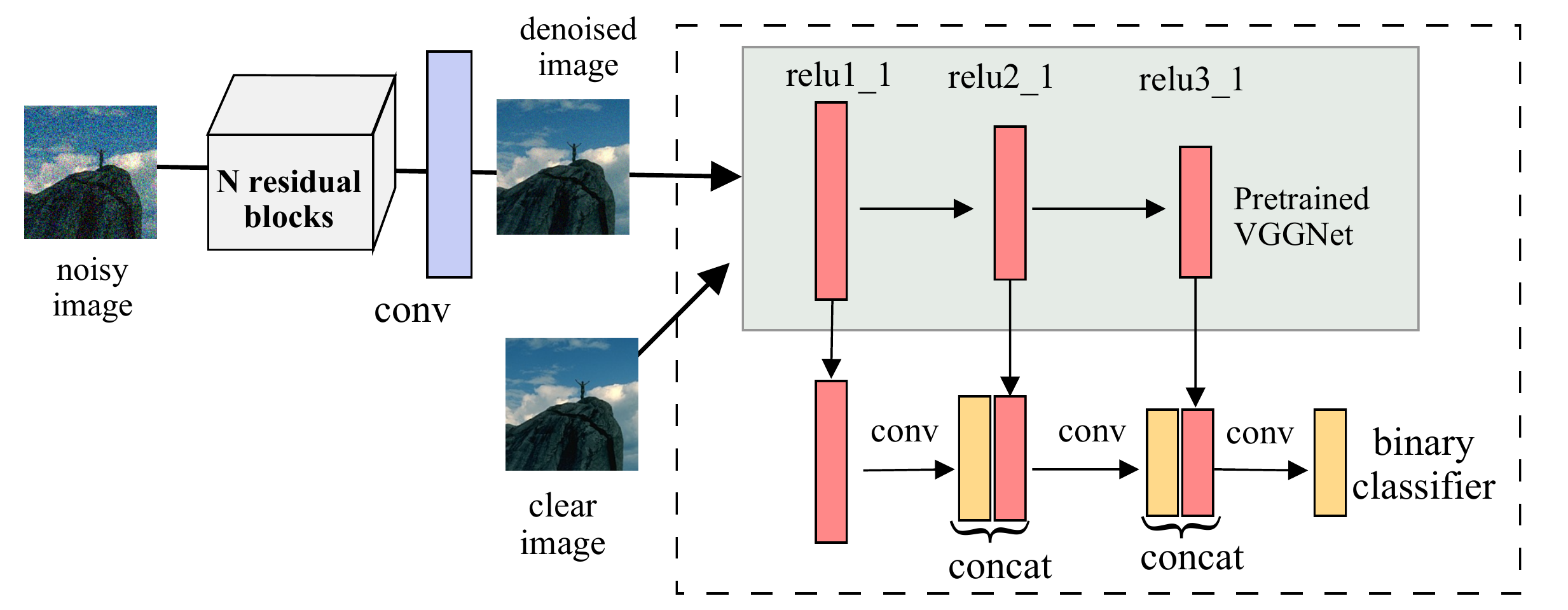}
\end{center}
   \caption{The dashed rectangle is used to construct the loss function to learn pixel-level prior.
   }
\label{fig:pixel_prior}
\end{figure}

\textbf{Pixel-level Prior.}
The pixel-level prior $P_{pix}$ is designed to push the denoised image to the natural image manifold to ensure realistic outputs. To achieve this, we employ a patch-based discriminator under the GAN framework. We adopt a \emph{perceptual discriminator} \cite{sungatullina2018image} to stabilize and improve the performance of GAN by embedding the convolutional parts of a pre-trained deep classification network (Figure \ref{fig:pixel_prior}). Specifically, the extracted features of the output image from the pre-trained network are concatenated with the output of the previous layer, and then processed by learnable blocks of convolutional operations. We use $3$ stride convolutional blocks to achieve spatial downsampling and use $relu1\_1$, $relu2\_1$ and $relu3\_1$ of VGG-19 \cite{simonyan2015very} for feature extraction. The final classification is processed on each activation from the feature map. As the effective receptive field for each activation corresponds to an image patch on the input image \cite{ren2015faster}, the discriminator actually predicts each label for each image patch. Patch based discriminator is quite useful to model high-frequencies in image denoising by restricting our attention to the structure in local image patches.

The optimization of discriminator is similar to classic binary classification. Specifically let us denote $D$ as the label of input image, thus we assign $D$ = 1 for denoised images and 0 for clear images; $p^{(w, h)}$ represents the feature map activation of the discriminator at location $(w, h)$. Then the pixel-level prior ($p_{pix}$) loss with $N$ samples can be written as:
\begin{scriptsize}
\begin{equation}
\label{eq:loss_binary}
\mathcal{L}_{p_{pix}} = - \frac{1}{N} \sum_{i=1}^N D_i log(p_i^{(w, h)}) + (1 - D_i) log(1-p_i^{(w, h)})
\end{equation}
\end{scriptsize}
As discussed in Section \ref{sec:hdivergence}, we try to simultaneously minimize above loss with respect to the discriminator and maximize it with respect to the transformation network. This can be achieved by the training strategy of generative adversarial network.

\subsection{Transformation Network}
\label{sec:network}
\begin{figure*}
\begin{center}
\includegraphics[width=18cm]{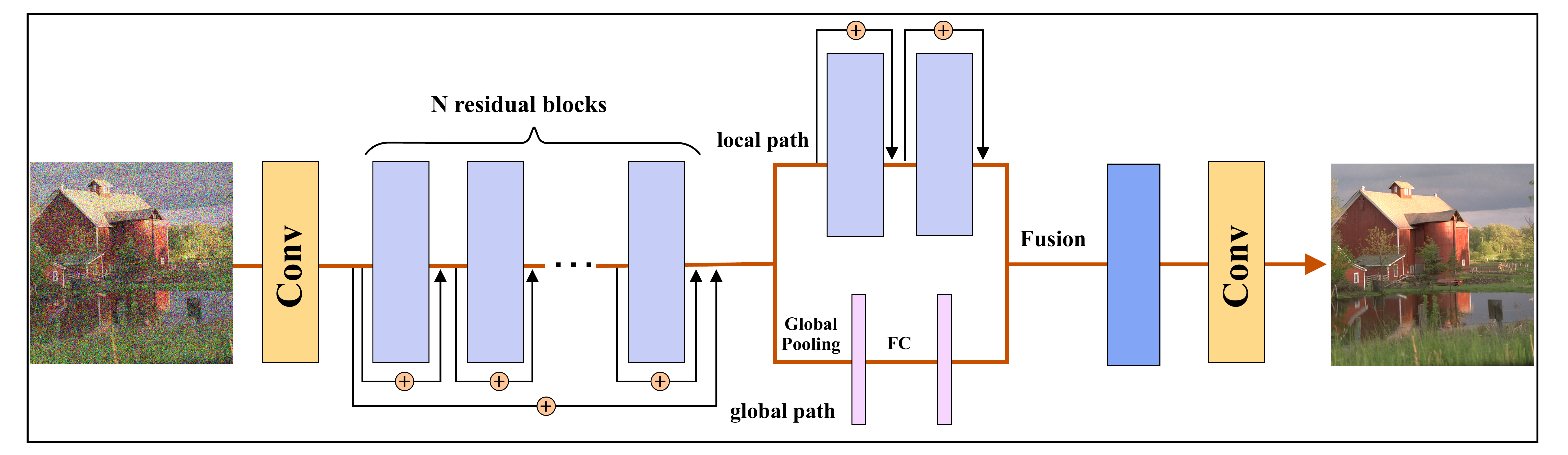}
\end{center} 
   \caption{Overview of our transformation network. The input noisy image is first processed by a deep residual network to compute low-level features, which are further split into two paths to learn both local and global features. Our model then fuses these two paths to produce the final output.
   }
\label{fig:overview}
\end{figure*}

Inspired by the architectural guidelines of several previous works \cite{he2016deep, ledig2017photo,johnson2016perceptual}, our image transformation network consists of 3 components: a stack of residual blocks (Figure \ref{fig:overview}) to extract low-level features of the input image and two asymmetric paths to extract local and global features respectively. Our architecture then fuses these two paths to produce the final output.

\textbf{Low-level Path.}
The input noisy image is first processed by a 16-layer residual network with skip-connection to extract low-level features (Figure \ref{fig:overview}). The ``pre-activation'' residual block \cite{he2016identity} is adopted  as it is much easier to train and generalizes better than the original ResNet \cite{he2016deep}. For all the residual blocks, a kernel size of $3 \times 3$ and zero-padding are used to keep the spatial size of the input. We also keep the number of features constant as 32 in all the residual blocks. Moreover, a skip-connection is added between the input features and the output of the last residual block. As a result, complex patterns can be extracted with a large spatial support.

\textbf{Local Path and Global Path.}
Similar to \cite{gharbi2017deep,IizukaSIGGRAPH2016}, the encoded features are further processed by two asymmetric networks for local and global feature extraction.

The \emph{local path} is fully convolutional and consists of two residual blocks as shown in Figure \ref{fig:overview}. It specializes in learning local features while retaining the spatial information. As argued in \cite{he2016deep}, the residual connections make it easy to learn identical function, which is an appealing property for the transformation network considering that the output image shares a lot of structures with the input image.

The \emph{global path} uses two fully-connected layers to learn global features. Each fully-connected layer is followed by a ReLU layer as the activation function. A global average pooling layer \cite{lin2013network} is used to ensure that our model can process images of any resolution. Finally, the global information is summarized as a fixed dimensional vector and used to regularize the local features produced by the local path.

The local and global features are then \emph{fused} into a common set of features, which are fed to a convolutional layer to produce the output. The fusion is achieved by a point-wise affine mixing with ReLU non-linearity as follows:

\begin{scriptsize}
\begin{equation}
F_c[x, y] = ReLU \Big(\sum_{c'} w'_{cc'}G_{c'} + \sum_{c'}w_{cc'}L_{c'}[x, y] + b_c \Big)
\end{equation}
\end{scriptsize}
where the $F_c[x, y]$ is the fused activation at the point $[x, y]$ of $c^{th}$ channel. $G_{c'}$ and $L_{c'}$ denote the global and local features. $c$ and $c'$ are the number of channels of fused and input features. $w'_{cc'}$ and $w_{cc'}$ are learnable weights to compute point-wise combinations of global and local features. This can be implemented with common convolutional operation with $1 \times 1$ kernels, which yields a fused 3-d array as the same shape as the input local features.

\subsection{Loss Function}
Pixel-wise mean square error (MSE) between two images is widely used as an optimization target for image denoising problem. However, MSE could result in unsatisfying results with over-smooth textures. In this work, we adopt $L_1$ loss as it encourages less blurring for image reconstruction \cite{isola2017image}. Specially the $L_1$ loss between denoised image $\hat{x}$ and clear image $x$ of shape [$C$, $W$, $H$] is calculated as:
\begin{equation}
\mathcal{L}_1 =  \frac{1}{CWH} \sum_{c=1}^{C} \sum_{w=1}^{W}  \sum_{h=1}^{H} \|x_{c,w,h} - \hat{x}_{c,w,h}\|
\end{equation}
Therefore the final training losses for feature-level and pixel-level prior of the proposed method are as follows:

\begin{equation}
\mathcal{L}_{feat} = \mathcal{L}_1 + \lambda_1\mathcal{L}_{p_{feat}}
\end{equation}
\begin{equation}
\mathcal{L}_{pix} = \mathcal{L}_1 +  \lambda_2\mathcal{L}_{p_{pix}}
\end{equation}
In our experiments, the hyper-parameter $\lambda_1$ and $\lambda_2$ is fixed to 0.001 to balance the fidelity loss and the image prior loss. From Figure \ref{fig:feat_prior} and \ref{fig:pixel_prior}, we can see that all the components can be trained jointly in an end-to-end manner using a standard SGD algorithm.

\begin{figure*}
\begin{center}
\includegraphics[width=17cm]{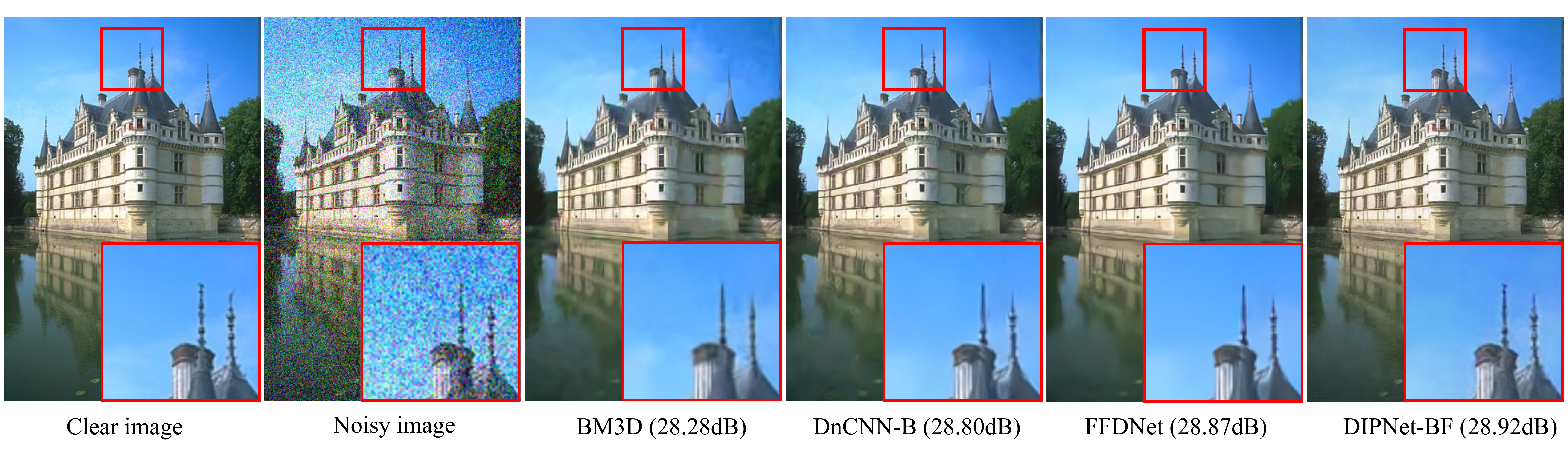}
\end{center}
   \caption{Denoising comparisons of an image from CBSD68 dataset with noise level $\sigma$ = 50.
   }
\label{fig:color_denoise}
\end{figure*}

\section{Experiments}
We conduct experiments on synthetic images for additive white Gaussian noise removal with either known or unknown noise levels as well as real-world noisy image denoising. 

\subsection{Datasets and Experiment Setting}
We train our models on the 2014 Microsoft COCO dataset \cite{lin2014microsoft}, which is a large-scale dataset containing 82,783 training images. We randomly crop image patches of size 64 $\times$ 64 for training. The input noisy images are obtained by adding AWGN of noise level $\sigma \in [15, 75]$ while the corresponding clear images are used as ground-truth. The noisy images are not clipped as previous works \cite{zhang2017beyond,zhang2018ffdnet}. To evaluate our denoiser on Gaussian noise removal, we use 3 datasets, \ie, CBSD68 \cite{roth2005fields} (68 images), Kodak24\cite{franzen1999kodak} (24 images) McMaster \cite{zhang2011color} (18 image) with synthetic noise. In addition, two benchmarks are considered to evaluate our method for real-world noise removal. The dataset1 is provided in \cite{xu2018real}, which includes 100 cropped images for evaluation and dataset2 is provided by in \cite{nam2016holistic}, which includes noisy images of 11 static scenes. Both datasets provide a mean image as the ``ground truth'', with which PSNR and SSIM can be calculated.

Our DIPNet employs 3 $\times$ 3 kernel for all the convolutional layers in the transformation network and the two discriminator networks. Each convolutional layer is followed by a batch normalization layer \cite{ioffe2015batch} to stabilize and accelerate the deep network training. We train all the models using Adam \cite{kingma2015adam} to achieve stochastic optimization with a batch size of 64 for 30 epochs and it takes around 24 hours for our model to get converged. The initial learning rate is $10^{-3}$, which is smoothly annealed by the cosine shape learning rate schedule introduced by \cite{loshchilov2017sgdr}. Our implementation is based on deep learning framework Pytorch and a single GTX 1080Ti GPU.



\subsection{AWGN noise Removal}
\label{sec:results}
\textbf{Non-blind AWGN Removal.}
We first test our DIPNet-S on noisy images corrupted with a specific noise level $\sigma$, referring non-blind AWGN removal. In other words, we train separate models for different noise levels without any image priors. We compare our method with CBM3D \cite{dabov2007image} and FFDNet \cite{zhang2018ffdnet}. Table \ref{tab:psnr_color} reports denoising results on different datasets. We can see that our DIPNet-S can achieve state of the art results and outperform other methods. Moreover, the improvement generalizes well across different datasets as well as different noise levels.

\begin{table}[!tb]
\begin{center}
\footnotesize
\tabcolsep=0.18cm
\begin{tabular}{|c|c|c|c|c|c|c|}
\hline
Dataset   & Method          & $\sigma$=15    & $\sigma$=25    & $\sigma$=35    & $\sigma$=50    & $\sigma$=75    \\
\hline\hline
 \multirow{3}{*}{CBSD68}         & CBM3D      & 33.52 & 30.71 & 28.89 & 27.38 & 25.74 \\
         & FFDNet     & 33.87 & 31.21 & 29.58 & \textbf{27.92} & \textbf{26.24} \\
         & DIPNet-S   & \textbf{33.90} & \textbf{31.25} & \textbf{29.60} & 27.91 & 26.16 \\

\hline
\multirow{3}{*}{Kodak24}         & CBM3D      & 34.28 & 31.68 & 29.90  & 28.46 & 26.82 \\
         & FFDNet     & 34.63 & 32.13 & \textbf{30.57} & \textbf{28.98} & \textbf{27.27} \\
         & DIPNet-S   & \textbf{34.64} & \textbf{32.15} & 30.56 & 28.97 & \textbf{27.27} \\
\hline
\multirow{3}{*}{McMaster}        & CBM3D      & 34.06 & 31.66 & 29.92 & 28.51 & 26.79 \\
       & FFDNet     & 34.66 & \textbf{32.35} & 30.81 & 29.18 & 27.33 \\
         & DIPNet-S   & \textbf{34.67} & \textbf{32.35} & \textbf{30.88} & \textbf{29.19} & \textbf{27.35} \\
\hline
\end{tabular}
\end{center}
\caption{Non-blind denoising results of different methods on CBSD68, Kodak24 and McMaster for AWGN noise with level $\sigma=$ 15, 25, 35, 50 and 75.}
\label{tab:psnr_color}
\end{table}

\begin{table}[!tb]
\begin{center}
\footnotesize
\tabcolsep=0.1cm
\begin{tabular}{|c|c|c|c|c|c|c|c|}
\hline
Dataset                    & Method    & $\sigma$=15 & $\sigma$=25 & $\sigma$=35 & $\sigma$=50  & $\sigma$=75  & average                   \\
\hline\hline

\multirow{2}{*}{CBSD68}    & DnCNN-B    & 33.89       & 31.23       & 29.58       & 27.92 &                        24.47   & 29.42                        \\
                           & DIPNet-B   & 33.88 & 31.16 & 29.49 & 27.89 & 26.01 & 29.69\\
                     & DIPNet-BP  & 33.70 & 31.07 &   29.36 &  27.82 & 25.99 & 29.59 \\
                     & DIPNet-BF  & 33.86 & 31.24 & 29.59 & 27.93 & 26.29 & \textbf{29.78} \\
\hline
\multirow{2}{*}{Kodak24}  &  DnCNN-B    & 34.48       & 32.03       & 30.46       & 28.85 &25.04   & 30.17                        \\
                           & DIPNet-B   & 34.44 & 32.01 & 30.47 & 28.90 & 27.19 & 30.60\\
                     & DIPNet-BP  & 33.49 & 31.18 &   29.75 &  28.20 & 26.32 & 29.76 \\
                           & DIPNet-BF & 34.62       & 32.11       & 30.55       & 28.97  & 27.28 & \textbf{30.71} \\ 
\hline
\multirow{2}{*}{McMaster}  & DnCNN-B    & 33.44       & 31.51       & 30.14       & 28.61 & 25.10   & 29.76                        \\
                           & DIPNet-B   & 34.33 & 32.09 & 30.61 & 28.76 & 25.86 & 30.33\\
                     & DIPNet-BP  & 33.54 & 31.52 & 30.08 & 28.14 & 25.48 & 29.75 \\
                           & DIPNet-BF & 34.56       & 32.33       & 30.56       & 29.18 & 27.32  & \textbf{30.79} \\
\hline                        
\end{tabular}
\end{center}
\caption{Blind denoising results of different methods on CBSD68, Kodak24 and McMaster for AWGN noise with level $\sigma=$ 15, 25, 35, 50 and 75.}
\label{tab:psnr_blind}
\end{table}

\textbf{Blind AWGN Removal.}
We further extend our model for blind image denoising with unknown noise levels. Unlike most previous works \cite{burger2012image,chen2017trainable} that need to first estimate the noise level and then select the denoising model trained with the corresponding noise level, we train three blind models in our experiments: DIPNet-B refers to the model trained without any image priors, DIPNet-BF and DIPNet-BP refer to the models trained with feature-level and pixel-level priors respectively. Specifically, 5 noise levels ($\sigma = 15, 25, 35, 50, 75$) are adopted to train our blind denoising models. We compare our method with state of the art method DnCNN-B on different datasets. As shown in table \ref{tab:psnr_blind}, our DIPNet-BF can achieve state of the art results for blind AWGN noise removal. DIPNet-BF is flexible enough to handle a wide range of noise levels effectively with a single network. In addition, we compare the visual results of different methods for an image in CBSD68 corrupted with noise level $\sigma = 50$. From Figure \ref{fig:color_denoise}, we can see that BM3D shows slightly blurred results, DnCNN-B and FFDNet could produce over-smooth edges and textures. Instead, DIPNet-BF can produce the best perceptual quality of denoised images with sharp edges and fine details.
\begin{figure}[!tb]
\begin{center}
\includegraphics[width=0.5\textwidth]{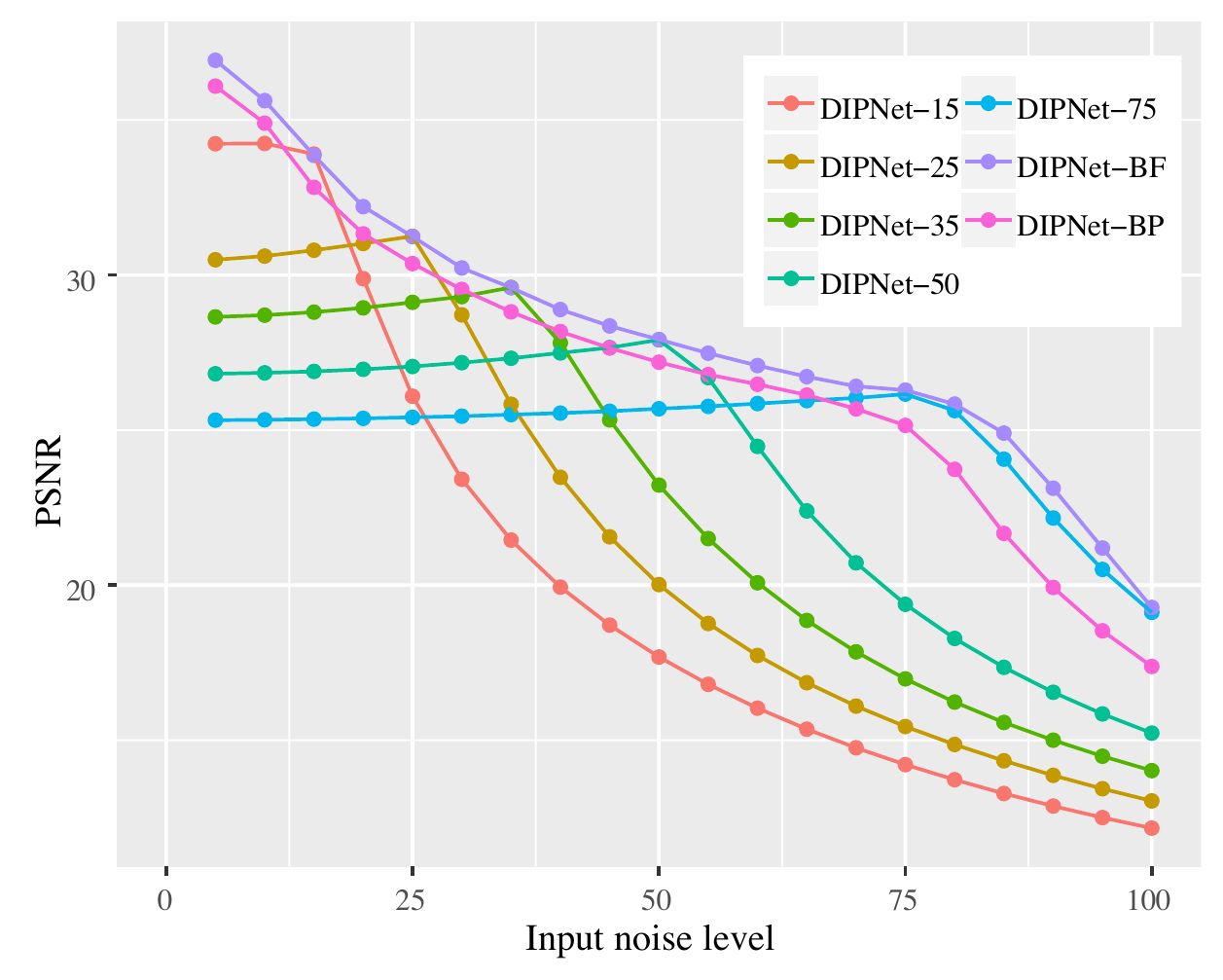}
\end{center}
   \caption{Noise level sensitivity curves of DIPNet models trained with different noise levels. The averaged PSNR results are evaluated on CBSD68 for different input noise levels.
   }
\label{fig:noise_sensitive}
\end{figure}

\begin{table*}[!tb]
\begin{center}
\scriptsize
\begin{tabular}{|c|c|c|c|c|c|c|c|c|c|c|c|}
\hline
Dataset                                         & Metric & CBM3D  & WNNM & MLP & TNRD   & DnCNN  & NI  &Guided   & DIPNet-B & DIPNet-BF & DIPNet-BP \\ \hline\hline
\multicolumn{1}{|c|}{\multirow{2}{*}{Dataset1\cite{xu2018real}}} & PSNR   & 37.40  & 36.59 & 38.07  & 38.17  & 36.08  & 37.77 &38.35 & 38.44 & \textbf{38.47} & 37.89  \\ \cline{2-12} 
\multicolumn{1}{|c|}{}                          & SSIM   & 0.9526 & 0.9247 &0.9615 & 0.9640 & 0.9161 & 0.9570 & 0.9669  & 0.9669 & \textbf{0.9676} & 0.9636 \\ \hline
\multirow{2}{*}{Dataset2\cite{nam2016holistic}}                       & PSNR   & 35.19  & 35.77 &36.46  & 36.61  & 33.86  & 35.49 &37.15  & 37.23 &\textbf{37.45} & 35.94 \\ \cline{2-12} 
                                                & SSIM   & 0.8580 & 0.9381 &0.9436 & 0.9463 & 0.8635 & 0.9126 &\textbf{0.9504} & 0.9389 &0.9452 & 0.9217 \\ \hline
\end{tabular}
\end{center}
\caption{Average PSNR(dB) and SSIM of different denoising methods on real-world noisy images in dataset1 \cite{xu2018real} and dataset2 \cite{nam2016holistic}.}
\label{tab:psnr_real}
\end{table*}

\subsection{Noise Level Sensitivity}
We further conduct experiments to investigate the noise level sensitivity of the proposed DIPNet models considering that the input noise level is usually unknown or very hard to estimate in practice. We compare the denoising performances of several different non-blind and blind DIPNet models with different input noise levels. Figure \ref{fig:noise_sensitive} shows the noise level sensitivity curves of different DIPNet models. Specifically, we consider the 5 non-blind DIPNets trained with known noise levels, \eg, ``DIPNet-15'' represents DIPNet trained with the fixed noise level $\sigma = 15$. We also compare the results of DIPNet-BF and DIPNet-BP. As shown in Figure \ref{fig:noise_sensitive}, we have the following two observations:

\begin{itemize}
   \item For non-blind DIPNets, the best performances are achieved when the input noise level matches the noise level used for training. The PSNR values decrease slowly when using lower input noise levels and begin to drops significantly when the input noise levels surpass the training noise levels.

   \item Our DIPNet-BF and DIPNet-BP demonstrate more stable performance with different input noise levels and the PSNR values denoising slowly with higher input noise levels. DIPNet-BF is capable of generalizing well to a wide range noise level (5, 100) when trained only on 5 fixed noise levels and outperforms nearly all the non-blind models for different input noise levels.

\end{itemize}

Based on the above observations, it is clear that the non-blind DIPNet-S models with specific noise levels are more sensitive to input noise, especially higher levels. DIPNet-BF demonstrates much stable performance in terms of a wide range of noise levels. DIPNet-BP is more sensitive to higher level noise.

\begin{figure*}[!tb]
\begin{center}
\includegraphics[width=17cm]{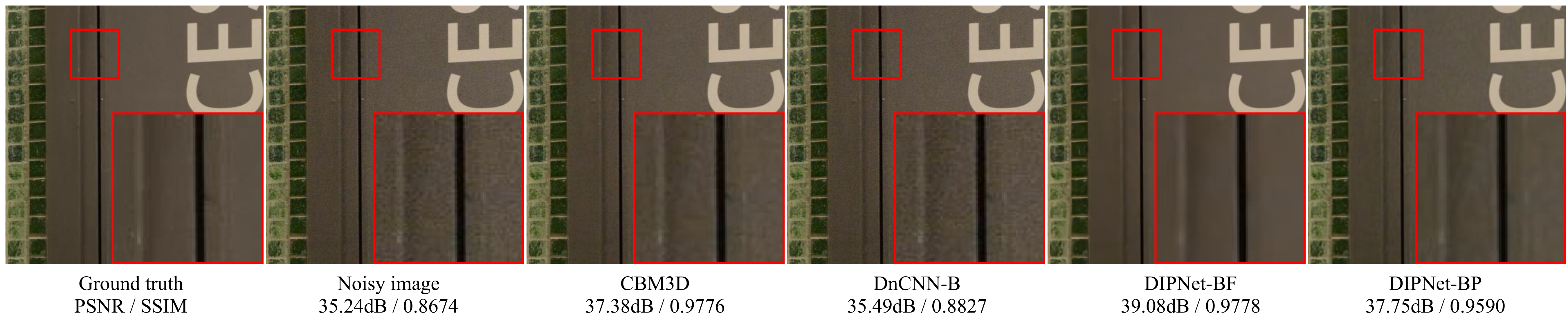}
\end{center}
   \caption{Denoising results on a real noisy image with different method. The test real images are from dataset1 \cite{xu2018real}.
   }
\label{fig:real_noise}
\end{figure*}

\begin{figure*}[!tb]
\begin{center}
\includegraphics[width=17cm]{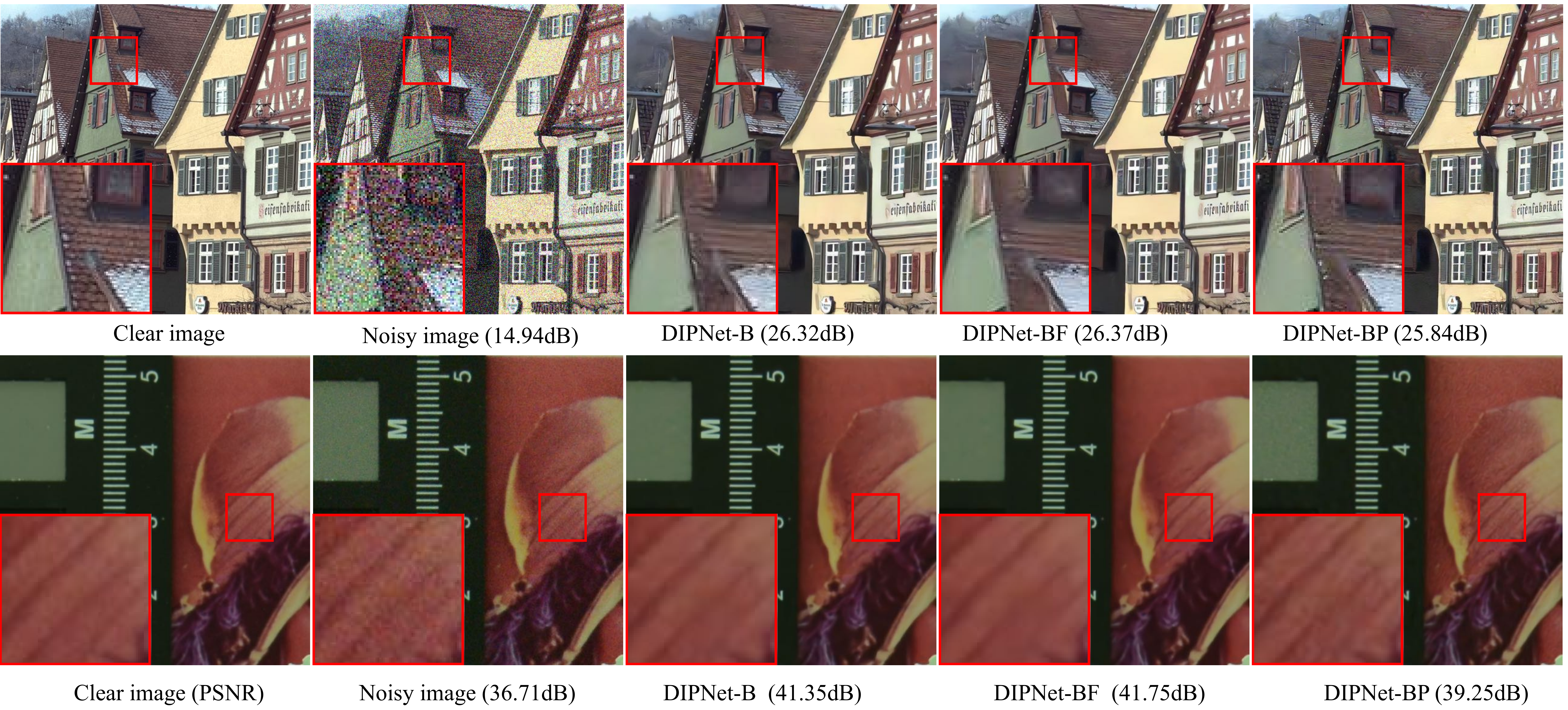}
\end{center}
   \caption{Noise removal comparisons of our methods. The first row is the results of AWGN noise removal for an image from Kodak24 dataset with noise level $\sigma$ = 50. The second row is the results of real noise removal for an image from dataset2 \cite{nam2016holistic}.
   }
\label{fig:gan_denoise}
\end{figure*}



\subsection{Real Noise Removal}
Furthermore, we evaluate our blind models on real noisy image datasets provided by \cite{nam2016holistic} and \cite{xu2018real} in terms of PSNR and SSIM. We compare with CBM3D \cite{dabov2007image}, DnCNN \cite{zhang2017beyond}, WNNM\cite{gu2014weighted}, TNRD\cite{chen2015learning} and MLP \cite{burger2012image}, which are state-of-the-art methods for AWGN noise removal. We also compare with Neat Image (NI) which is a set of commercial software for image denoising \cite{xu2018real} and recent Guided \cite{xu2018external}. Table \ref{tab:psnr_real} reports denoising results on the two datasets. We can see that our model trained with feature-level prior (DIPNet-BF) can achieve superior performance compared with traditional methods like CBM3D and deep learning based DnCNN. Moreover, our method can outperform Guided algorithm\cite{xu2018external}, which is designed specifically for realistic noise removal. We provide a visual example by different denoising methods from dataset1 \cite{xu2018real} in Figure \ref{fig:real_noise}. We can observe that using feature-level prior can effectively improve PSNR but still produce over-smooth textures. The pixel-level prior can help produce sharper appearance but with lower PSNR and SSIM. This is due to the appearance of high-frequency artifacts.

\subsection{Effectiveness of Image Priors}
We further demonstrate the effectiveness of the two image priors in our model. We consider blind image denoising trained with 5 noise levels ($\sigma = 15, 25, 35, 50, 75$) in Section \ref{sec:results}. Quantitative results for AWGN and real noise removal are summarized in Table \ref{tab:psnr_blind} and Table \ref{tab:psnr_real}. Two visual examples are provided in Figure \ref{fig:gan_denoise}. The first row shows AWGN noise removal results for an image from Kodak24, the second row shows the real noise removal for an image from dataset2 \cite{nam2016holistic}. We can see that DIPNet-BF trained with feature-level prior provides the highest PSNR values for different noise levels, especially for higher noise levels. What's more, DIPNet-BF shows superior performance over DIPNet-B on real noise removal task. It is clear that the feature-level prior can effectively improve the generalizability when adapting from synthetic Gaussian denoising to real noise removal by encoding domain-invariant features. On the other hand, incorporating pixel-level prior degrades the denoising performance of DIPNet in terms of PSNR. However, we observe that using pixel-level prior can yield better texture details and sharp edges when compared to clear images (Figure \ref{fig:gan_denoise}) while feature-level prior could produce slightly over-smooth images. The pixel-level prior can achieve perceptual improvements to produce photo-realistic images but with lower PSNR values by introducing high-frequency artifacts.

\begin{table}[!tb]
\begin{center}
\footnotesize
\begin{tabular}{|c|c|c|}
\hline
dataset  & DIPNet-BF & DIPNet-BP \\
\hline
\hline
CBSD68   & 0.3404    & \textbf{0.6596}    \\
dataset1 & 0.3857    & \textbf{0.6143}   \\
\hline
\end{tabular}
\end{center}
\caption{Subjective evaluation results on CBSD68 \cite{roth2005fields} and dataset1 \cite{xu2018real}. We show the average preference for all the volunteers.}
\label{tab:mos}
\end{table}

\subsection{Subjective Evaluation}
We have conducted a subjective evaluation to quantify the ability of the two image priors to produce perceptually convincing images. In particular, we show volunteers 4 images each time, i.e., the noisy image, ground-truth image and two denoised images produced by DIPNet-BF and DIPNet-BP and ask them which denoised version they prefer. We ask 12 volunteers to performance the evaluation on CBSD68 (68 images) \cite{roth2005fields} and dataset1 (100 images) \cite{xu2018real} for both synthetic and real noise removal respectively. The experimental results of the voting test across all the volunteers are summarized in Table \ref{tab:mos}. We can see that DIPNet-BP outperforms DIPNet-BF by a large margin and DIPNet-BP is superior in terms of perceptual quality. It is because pixel-level prior can produce high-frequency details as shown in Figure \ref{fig:gan_denoise}, which are visually pleasing especially for images with high texture details. However, the generated high-frequency details fail to exactly match the fidelity expected in the clear images. As a result, the denoising performances with pixel-level prior is inferior in terms of distortion measures like PSNR and SSIM. Additional examples are depicted in the supplementary material.

\section{Conclusion}
In this paper, we have developed an effective blind image denoising model based on data-driven image priors. The two image priors are designed from the perspective of domain alignment. Specifically the feature-level prior can help alleviate the domain discrepancy across different level noise and improve the blind image denoising performance. The pixel-level prior is able to push the denoised outputs to the natural image manifold for perceptual quality improvement. This is further confirmed by a subjective evaluation. The two image priors are learned based on adversarial training of $\mathcal{H}$-divergence using the standard SGD optimization technique. Validated on various datasets, our approach can achieve state of the art results for both synthetic and real-world noise removal.

{\small
\bibliographystyle{ieee}
\bibliography{egbib}

\begin{thebibliography}{10}\itemsep=-1pt

\bibitem{agostinelli2013adaptive}
Forest Agostinelli, Michael~R Anderson, and Honglak Lee.
\newblock Adaptive multi-column deep neural networks with application to robust
  image denoising.
\newblock In {\em NIPS}, 2013.

\bibitem{ben2010theory}
Shai Ben-David, John Blitzer, Koby Crammer, Alex Kulesza, Fernando Pereira, and
  Jennifer~Wortman Vaughan.
\newblock A theory of learning from different domains.
\newblock {\em Machine learning}, 79(1-2):151--175, 2010.

\bibitem{ben2007analysis}
Shai Ben-David, John Blitzer, Koby Crammer, and Fernando Pereira.
\newblock Analysis of representations for domain adaptation.
\newblock In {\em NIPS}, 2007.

\bibitem{blau2018perception}
Yochai Blau and Tomer Michaeli.
\newblock The perception-distortion tradeoff.
\newblock In {\em CVPR}, 2018.

\bibitem{buades2005non}
Antoni Buades, Bartomeu Coll, and J-M Morel.
\newblock A non-local algorithm for image denoising.
\newblock In {\em CVPR}, 2005.

\bibitem{burger2012image}
Harold~C Burger, Christian~J Schuler, and Stefan Harmeling.
\newblock Image denoising: Can plain neural networks compete with bm3d?
\newblock In {\em CVPR}, 2012.

\bibitem{chen2018deep}
Chang Chen, Zhiwei Xiong, Xinmei Tian, and Feng Wu.
\newblock Deep boosting for image denoising.
\newblock In {\em ECCV}, 2018.

\bibitem{chen2018image}
Jingwen Chen, Jiawei Chen, Hongyang Chao, and Ming Yang.
\newblock Image blind denoising with generative adversarial network based noise
  modeling.
\newblock In {\em CVPR}, 2018.

\bibitem{chen2018domain}
Yuhua Chen, Wen Li, Christos Sakaridis, Dengxin Dai, and Luc Van~Gool.
\newblock Domain adaptive faster r-cnn for object detection in the wild.
\newblock In {\em CVPR}, 2018.

\bibitem{chen2017trainable}
Yunjin Chen and Thomas Pock.
\newblock Trainable nonlinear reaction diffusion: A flexible framework for fast
  and effective image restoration.
\newblock {\em IEEE TPAMI}, 39(6):1256--1272, 2017.

\bibitem{chen2015learning}
Yunjin Chen, Wei Yu, and Thomas Pock.
\newblock On learning optimized reaction diffusion processes for effective
  image restoration.
\newblock In {\em CVPR}, 2015.

\bibitem{dabov2007image}
Kostadin Dabov, Alessandro Foi, Vladimir Katkovnik, and Karen Egiazarian.
\newblock Image denoising by sparse 3-d transform-domain collaborative
  filtering.
\newblock {\em IEEE TIP}, 16(8), 2007.

\bibitem{dong2016image}
Chao Dong, Chen~Change Loy, Kaiming He, and Xiaoou Tang.
\newblock Image super-resolution using deep convolutional networks.
\newblock {\em IEEE TPAMI}, 38(2):295--307, 2016.

\bibitem{Dong2013NonlocallyCS}
Weisheng Dong, Lei Zhang, Guangming Shi, and Xin Li.
\newblock Nonlocally centralized sparse representation for image restoration.
\newblock {\em IEEE TIP}, 22:1620--1630, 2013.

\bibitem{elad2006image}
Michael Elad and Michal Aharon.
\newblock Image denoising via sparse and redundant representations over learned
  dictionaries.
\newblock {\em IEEE TIP}, 15(12):3736--3745, 2006.

\bibitem{franzen1999kodak}
Rich Franzen.
\newblock Kodak lossless true color image suite.
\newblock {\em source: http://r0k. us/graphics/kodak}, 4, 1999.

\bibitem{ganin2015unsupervised}
Yaroslav {Ganin} and Victor~S. {Lempitsky}.
\newblock Unsupervised domain adaptation by backpropagation.
\newblock {\em ICML}, 2015.

\bibitem{gharbi2017deep}
Michaël {Gharbi}, Jiawen {Chen}, Jonathan~T. {Barron}, Samuel~W. {Hasinoff},
  and Frédo {Durand}.
\newblock Deep bilateral learning for real-time image enhancement.
\newblock {\em ACM Transactions on Graphics}, 36(4):1--12, 2017.

\bibitem{goodfellow2014generative}
Ian Goodfellow, Jean Pouget-Abadie, Mehdi Mirza, Bing Xu, David Warde-Farley,
  Sherjil Ozair, Aaron Courville, and Yoshua Bengio.
\newblock Generative adversarial nets.
\newblock In {\em NIPS}, 2014.

\bibitem{gu2014weighted}
Shuhang Gu, Lei Zhang, Wangmeng Zuo, and Xiangchu Feng.
\newblock Weighted nuclear norm minimization with application to image
  denoising.
\newblock In {\em CVPR}, 2014.

\bibitem{he2016deep}
Kaiming He, Xiangyu Zhang, Shaoqing Ren, and Jian Sun.
\newblock Deep residual learning for image recognition.
\newblock In {\em CVPR}, 2016.

\bibitem{he2016identity}
Kaiming He, Xiangyu Zhang, Shaoqing Ren, and Jian Sun.
\newblock Identity mappings in deep residual networks.
\newblock In {\em ECCV}, 2016.

\bibitem{heide2014flexisp}
Felix Heide, Markus Steinberger, Yun-Ta Tsai, Mushfiqur Rouf, Dawid Pajak,
  Dikpal Reddy, Orazio Gallo, Jing Liu, Wolfgang Heidrich, Karen Egiazarian,
  et~al.
\newblock Flexisp: A flexible camera image processing framework.
\newblock {\em ACM Transactions on Graphics (TOG)}, 33(6):231, 2014.

\bibitem{hou2017deep}
Xianxu Hou, Jiang Duan, and Guoping Qiu.
\newblock Deep feature consistent deep image transformations: downscaling,
  decolorization and hdr tone mapping.
\newblock {\em arXiv preprint arXiv:1707.09482}, 2017.

\bibitem{hou2017deepvae}
Xianxu Hou, Linlin Shen, Ke Sun, and Guoping Qiu.
\newblock Deep feature consistent variational autoencoder.
\newblock In {\em WACV}, 2017.

\bibitem{IizukaSIGGRAPH2016}
Satoshi Iizuka, Edgar Simo-Serra, and Hiroshi Ishikawa.
\newblock {Let there be Color!: Joint End-to-end Learning of Global and Local
  Image Priors for Automatic Image Colorization with Simultaneous
  Classification}.
\newblock {\em ACM Transactions on Graphics}, 35(4), 2016.

\bibitem{ioffe2015batch}
Sergey {Ioffe} and Christian {Szegedy}.
\newblock Batch normalization: Accelerating deep network training by reducing
  internal covariate shift.
\newblock {\em ICML}, 2015.

\bibitem{isola2017image}
Phillip {Isola}, Jun-Yan {Zhu}, Tinghui {Zhou}, and Alexei~A. {Efros}.
\newblock Image-to-image translation with conditional adversarial networks.
\newblock In {\em CVPR}, 2017.

\bibitem{jain2009natural}
Viren Jain and Sebastian Seung.
\newblock Natural image denoising with convolutional networks.
\newblock In {\em NIPS}, 2009.

\bibitem{johnson2016perceptual}
Justin Johnson, Alexandre Alahi, and Li Fei-Fei.
\newblock Perceptual losses for real-time style transfer and super-resolution.
\newblock In {\em ECCV}, 2016.

\bibitem{kingma2015adam}
Diederik~P. {Kingma} and Jimmy~Lei {Ba}.
\newblock Adam: A method for stochastic optimization.
\newblock {\em ICLR}, 2015.

\bibitem{larsson2016learning}
Gustav Larsson, Michael Maire, and Gregory Shakhnarovich.
\newblock Learning representations for automatic colorization.
\newblock In {\em ECCV}, 2016.

\bibitem{ledig2017photo}
Christian Ledig, Lucas Theis, Ferenc Husz{\'a}r, Jose Caballero, Andrew
  Cunningham, Alejandro Acosta, Andrew~P Aitken, Alykhan Tejani, Johannes Totz,
  Zehan Wang, et~al.
\newblock Photo-realistic single image super-resolution using a generative
  adversarial network.
\newblock In {\em CVPR}, 2017.

\bibitem{lefkimmiatis2017non}
Stamatios Lefkimmiatis.
\newblock Non-local color image denoising with convolutional neural networks.
\newblock In {\em CVPR}, 2017.

\bibitem{li2018learning}
Lerenhan Li, Jinshan Pan, Wei-Sheng Lai, Changxin Gao, Nong Sang, and
  Ming-Hsuan Yang.
\newblock Learning a discriminative prior for blind image deblurring.
\newblock In {\em CVPR}, 2018.

\bibitem{lin2013network}
Min Lin, Qiang Chen, and Shuicheng Yan.
\newblock Network in network.
\newblock {\em arXiv preprint arXiv:1312.4400}, 2013.

\bibitem{lin2014microsoft}
Tsung-Yi Lin, Michael Maire, Serge Belongie, James Hays, Pietro Perona, Deva
  Ramanan, Piotr Doll{\'a}r, and C~Lawrence Zitnick.
\newblock Microsoft coco: Common objects in context.
\newblock In {\em ECCV}, 2014.

\bibitem{liu2008automatic}
Ce Liu, Richard Szeliski, Sing~Bing Kang, C~Lawrence Zitnick, and William~T
  Freeman.
\newblock Automatic estimation and removal of noise from a single image.
\newblock {\em IEEE TPAMI}, 30(2):299--314, 2008.

\bibitem{long2015fully}
Jonathan Long, Evan Shelhamer, and Trevor Darrell.
\newblock Fully convolutional networks for semantic segmentation.
\newblock In {\em CVPR}, 2015.

\bibitem{loshchilov2017sgdr}
Ilya {Loshchilov} and Frank {Hutter}.
\newblock Sgdr: Stochastic gradient descent with warm restarts.
\newblock {\em ICLR}, 2017.

\bibitem{mairal2009non}
Julien Mairal, Francis Bach, Jean Ponce, Guillermo Sapiro, and Andrew
  Zisserman.
\newblock Non-local sparse models for image restoration.
\newblock In {\em CVPR}, 2009.

\bibitem{mildenhall2018burst}
Ben Mildenhall, Jonathan~T Barron, Jiawen Chen, Dillon Sharlet, Ren Ng, and
  Robert Carroll.
\newblock Burst denoising with kernel prediction networks.
\newblock In {\em CVPR}, 2018.

\bibitem{nam2016holistic}
Seonghyeon Nam, Youngbae Hwang, Yasuyuki Matsushita, and Seon Joo~Kim.
\newblock A holistic approach to cross-channel image noise modeling and its
  application to image denoising.
\newblock In {\em CVPR}, pages 1683--1691, 2016.

\bibitem{rabie2005robust}
Tamer Rabie.
\newblock Robust estimation approach for blind denoising.
\newblock {\em IEEE TIP}, 14(11):1755--1765, 2005.

\bibitem{ren2015faster}
Shaoqing Ren, Kaiming He, Ross Girshick, and Jian Sun.
\newblock Faster r-cnn: Towards real-time object detection with region proposal
  networks.
\newblock In {\em NIPS}, 2015.

\bibitem{roth2005fields}
Stefan Roth and Michael~J Black.
\newblock Fields of experts: A framework for learning image priors.
\newblock In {\em CVPR}, 2005.

\bibitem{rudin1992nonlinear}
Leonid~I Rudin, Stanley Osher, and Emad Fatemi.
\newblock Nonlinear total variation based noise removal algorithms.
\newblock {\em Physica D: nonlinear phenomena}, 60(1-4):259--268, 1992.

\bibitem{santhanam2017generalized}
Venkataraman Santhanam, Vlad~I Morariu, and Larry~S Davis.
\newblock Generalized deep image to image regression.
\newblock In {\em CVPR}, 2017.

\bibitem{simonyan2015very}
Karen {Simonyan} and Andrew {Zisserman}.
\newblock Very deep convolutional networks for large-scale image recognition.
\newblock {\em ICLR}, 2015.

\bibitem{sungatullina2018image}
Diana {Sungatullina}, Egor {Zakharov}, Dmitry {Ulyanov}, and Victor
  {Lempitsky}.
\newblock Image manipulation with perceptual discriminators.
\newblock In {\em ECCV}, pages 587--602, 2018.

\bibitem{ulyanov2018deep}
Dmitry Ulyanov, Andrea Vedaldi, and Victor Lempitsky.
\newblock Deep image prior.
\newblock In {\em CVPR}, pages 9446--9454, 2018.

\bibitem{weiss2007makes}
Yair Weiss and William~T Freeman.
\newblock What makes a good model of natural images?
\newblock In {\em CVPR}, 2007.

\bibitem{xie2012image}
Junyuan Xie, Linli Xu, and Enhong Chen.
\newblock Image denoising and inpainting with deep neural networks.
\newblock In {\em NIPS}, 2012.

\bibitem{xu2018real}
Jun Xu, Hui Li, Zhetong Liang, David Zhang, and Lei Zhang.
\newblock Real-world noisy image denoising: A new benchmark.
\newblock {\em arXiv preprint arXiv:1804.02603}, 2018.

\bibitem{xu2018external}
Jun Xu, Lei Zhang, and David Zhang.
\newblock External prior guided internal prior learning for real-world noisy
  image denoising.
\newblock {\em IEEE Transactions on Image Processing}, 27(6):2996--3010, 2018.

\bibitem{zhang2017beyond}
Kai Zhang, Wangmeng Zuo, Yunjin Chen, Deyu Meng, and Lei Zhang.
\newblock Beyond a gaussian denoiser: Residual learning of deep cnn for image
  denoising.
\newblock {\em IEEE TIP}, 26(7):3142--3155, 2017.

\bibitem{Zhang_2017_CVPR}
Kai Zhang, Wangmeng Zuo, Shuhang Gu, and Lei Zhang.
\newblock Learning deep cnn denoiser prior for image restoration.
\newblock In {\em CVPR}, 2017.

\bibitem{zhang2018ffdnet}
Kai Zhang, Wangmeng Zuo, and Lei Zhang.
\newblock Ffdnet: Toward a fast and flexible solution for cnn-based image
  denoising.
\newblock {\em IEEE TIP}, 27(9):4608--4622, 2018.

\bibitem{zhang2011color}
Lei Zhang, Xiaolin Wu, Antoni Buades, and Xin Li.
\newblock Color demosaicking by local directional interpolation and nonlocal
  adaptive thresholding.
\newblock {\em Journal of Electronic imaging}, 20(2):023016, 2011.

\bibitem{zhu2016noise}
Fengyuan Zhu, Guangyong Chen, and Pheng-Ann Heng.
\newblock From noise modeling to blind image denoising.
\newblock In {\em CVPR}, 2016.

\end{thebibliography}
}

\end{document}